\documentclass[12pt]{article}
\usepackage{graphics}
\usepackage{latexsym}

\newcommand{\be}{\begin{equation}}
\newcommand{\ee}{\end{equation}}
\newcommand{\bea}{\begin{eqnarray}}
\newcommand{\eea}{\end{eqnarray}}

\newcommand{\lngra}{\longrightarrow}

\newcommand{\psibar}{\overline{\psi}}

\newcommand{\hpart}{\stackrel{\leftrightarrow}{\partial}}

\newcommand{\mat}[3]{\left<{#1}\left|{#2}\right|{#3}\right>}

\newcommand{\xp}{{x^{\prime}}}

\newcommand{\ra}{\rightarrow}

\newcommand{\half}{\raisebox{1pt}{$\scriptstyle \frac{1}{2}$}}

\newcommand{\figpath}{.}

\newlength{\updownindent}
\newlength{\leftrightindent}
\begin{document}

\thispagestyle{empty}
\renewcommand{\thefootnote}{\alph{footnote}}

\begin{flushright} MC/TH01/02 \\ SWAT/285 \end{flushright}

\begin{center}
  {\Large \bf Factorisation in Deeply Virtual Compton}\\[5pt]
  {\Large \bf Scattering: Local OPE Formalism and}\\[5pt]
  {\Large \bf Structure Functions}\\
\vspace{0.2in}
  {\bf B. E. White}\\
\vspace{0.1in}
  {\it Department of Physics and Astronomy,\footnote{%
    Current address}} \\
  {\it University of Manchester,}\\
  {\it Oxford Road,}\\
  {\it Manchester M13 9PL, U.K.}\\[5pt]
  and\\[5pt]
  {\it Physics Department,}\\
  {\it University of Wales Swansea,}\\
  {\it Singleton Park,}\\
  {\it Swansea SA2 8PP, U.K.}\\[5pt]
  E-mail: {\tt ben@theory.ph.man.ac.uk}\\
\vspace{0.15in}
\end{center}

\begin{abstract}

We give a complete treatment of factorisation of Deeply Virtual Compton
Scattering~(DVCS) in the generalised Bjorken limit, using the local
Operator Product Expansion~(OPE). The method allows a
straightforward proof that, at leading twist, the DVCS amplitude factorises
into an integral over coefficient functions and Skewed Parton Distribution
Functions~(SPDFs). The integral is well defined for on-shell final state
photon if the Wilson coefficients
satisfy a certain factorisation condition, which we derive.
We also show that it enables a simple
proof that soft singularities either cancel out or, in the case where
the final state photon is on shell, are integrable. This confirms the
argument of Collins and Freund. Further, we repeat the
tree-level calculation of twist-three contributions to DVCS off a scalar
target, where factorisation was found to be violated. We propose a new
definition of the structure functions and calculate the coefficient functions,
which are such that factorisation works.

\end{abstract}

\vspace{0.3in}

\newpage
\setcounter{page}{1}

\renewcommand{\thefootnote}{\arabic{footnote}}
\setcounter{footnote}{0}


\section{Introduction}
\label{sec_intro}

In recent years, there has been much interest in the process
known as Deeply Virtual Compton Scattering~(DVCS)~\cite{Expt},
$\gamma^\ast p \ra \gamma p$. In the generalised Bjorken regime, where
the invariant mass of the virtual photon
$\gamma^\ast$ is very large compared to all other kinematic scales
except the $\gamma p$ centre-of-mass energy squared, it has been
shown~\cite{Rad,OJ,CF,BGR} that the amplitude for DVCS factorises
into a perturbatively calculable coefficient
function and a non-perturbative factor, the
Skewed Parton Distribution Function~(SPDF)~\cite{Ji3}.
This factorisation generalises that of the
forward Compton amplitude relevant to Deep Inelastic Scattering (DIS).
Thus, measurements of DVCS can be used to extract information on SPDFs.
One-loop perturbative calculations have been performed~\cite{OJ,Man,BM}
for some of the possible Lorentz structures which demonstrate factorisation.

Our principal purpose will be to provide a complete treatment
of the factorisation of the leading term
for the DVCS amplitude from the point of view of the
local Operator Product Expansion~(OPE). 
Previous proofs of factorisation have applied rather different methods.
In refs.~\cite{OJ,CF}, factorisation was considered by analysing the
mass singularities that appear when all small scales in the problem go
to zero. All collinear mass singularities can then be absorbed by the
SDPFs. However, soft singularities, where soft lines connect the
hard coefficient to the SPDF, were shown to make leading order contributions.
For factorisation not to be spoiled one would have to show that the soft
singularities are in fact integrable~\cite{CF}.
This was shown explicitly to be the case at the
one-loop level, given that the SPDFs are sufficiently smooth. For the
all-order proof~\cite{CF}, a general argument~\cite{Col} can be applied.
An important feature of the local OPE treatment of the present paper
is that one can show quite easily that all such soft singularities either
cancel out or, where we put the final-state photon on shell,
are integrable, given the same smoothness assumption for the
SPDF. Moreover, the local OPE provides a concise and elegant scheme in which
to perform calculations.

More recently, the effect of contributions from twist-3 SDPFs has been
considered~\cite{BM2,BMKS,Tw3}.
It has been shown that, even at tree-level, they
are such that the factorisation formula diverges. The second
purpose of this article is to revisit these calculations. We shall
find that a suitable redefinition of the structure function to which the
twist-3 SPDFs contribute will be sufficient to give them a factorisation
formula which does not diverge.

In order to demonstrate how factorisation works in the local OPE method,
we will first consider an analogue of DVCS in 
scalar field theory. This will enable a demonstration of the method without
a plethora of indices. The local OPE analysis is a straightforward
extension of that applied to DIS. In DVCS we must include total derivative
operators in the OPE,
which have non-vanishing off-forward matrix elements since
the final state proton possesses non-zero momentum relative to its initial
state. Furthermore,
the local OPE is an expansion in an unphysical region, as in DIS. We will
use a dispersion relation to reconstruct the full, factorised leading twist
amplitude. We shall find that dispersion relation integral is
only defined provided that the Wilson coefficients obey a factorisation
condition.

We then move on to consider the local OPE for
the Compton amplitude at tree level in QCD. The organisation of the OPE
into terms each obeying current conservation is demonstrated.
In particular, for the case of an unpolarised proton, we propose a
suitable basis of four structure functions, $T_T$, $T_L$ and $T^\pm_3$,
each multiplying a Lorentz structure which respects current conservation.
Since the OPE is an expansion of
two conserved currents, we shall check that the OPE is indeed consistent
with current conservation. However, as recently discussed in ref.~\cite{BM2},
the current conservation is only satisfied because of a cancellation of terms
between parity-even and -odd operators.
Fortunately, it turns out that the twist-three operators can, at tree level,
be related to the twist-2 ones~\cite{BM2}. Thus, the OPE
can be reorganised into two pieces of even and odd parity, each respecting
current conservation.  The twist-3 pieces contribute exclusively to the
$T^\pm_3$ structure functions. We calculate explicitly their coefficient
functions and show that factorisation is not violated. This is because
the corresponding Wilson coefficients obey the factorisation condition
mentioned above.

The article's contents are as follows: Section~\ref{sec_kin} briefly
describes the choices of kinematic variables used for DVCS;
section~\ref{sec_ope} contains the treatment of the local
OPE and dispersion relation used to reconstruct the factorised amplitude
from the local OPE, together with a simple model for the Wilson coefficients
from which are derived the corresponding coefficient functions;
in section~\ref{sec_qcd} we consider the OPE at tree level
in QCD and find the tree-level predictions for the four structure functions
in terms of SPDFs; finally we give a summary in section~\ref{sec_sum}. An
appendix on SPDFs is provided at the end.


\section{Kinematics}
\label{sec_kin}

\begin{figure}
\centering{\input{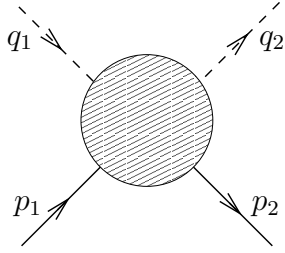}}
\caption{Definitions of momenta for DVCS.}
\label{fig_dvcs}
\end{figure}

In this section we give some definitions of kinematic variables.
In particular, we define the generalised Bjorken limit in which the
amplitude can be shown to factorise. Referring to fig.~\ref{fig_dvcs},
we label the initial and final state photons' momenta by $q_1$ and $q_2$
respectively and likewise the initial and final state proton's momenta
by $p_1$ and $p_2$. In terms of these, we define three independent
4-vectors:
\bea
  q & = & \frac{1}{2} \; (q_1 + q_2), \nonumber \\
  p & = & \frac{1}{2} \; (p_1 + p_2), \\
  \Delta & = & p_2 - p_1. \nonumber
\eea
The convenient, Lorentz-invariant kinematic variables are as follows:
\bea
  Q^2 & = & -q^2, \nonumber \\
  t & = & \Delta^2, \label{vars}\\
  x & = & \frac{Q^2}{2p\cdot q}, \nonumber \\
  \xi & = & \frac{q\cdot\Delta}{2q\cdot p}. \nonumber
\eea

The generalised Bjorken limit is defined by taking $Q^2$ large, while
keeping $x$ and $\xi$ finite. In this limit, we can neglect any quantities
suppressed by powers of $t/Q^2$ and $M^2/Q^2$, where $p_1^2 = p_2^2 = M^2$.

For future reference, we also write down the s- and u-channel variables
together with the invariant masses squared of the photons:
\bea
  s \; = \; (q + p)^2 & = & \frac{Q^2 \; (1-x)}{x} \; + \; M^2
   \; - \; \frac{1}{4} \; t, \nonumber \\
  u \; = \; (q - p)^2 & = & - \; \frac{Q^2 \; (1+x)}{x} \; + \; M^2
   \; - \; \frac{1}{4} \; t, \nonumber \\
  q^2_1 & = & - \; \frac{Q^2 \; (x - \xi)}{x} \; + \; \frac{1}{4} \; t,
   \label{invarts}\\
  q^2_2 & = & - \; \frac{Q^2 \; (x + \xi)}{x} \; + \; \frac{1}{4} \; t.
   \nonumber
\eea
Since $q^2_2 = 0$ for DVCS, then we have a constraint on the four, otherwise
independent variables in eqs.~(\ref{vars}):
\be
  \xi \; = \; - \, x \left( 1 \; + \; \frac{t}{4Q^2}\right).
\ee
However, for the purposes of the local OPE analysis in section~\ref{sec_ope},
we relax this constraint.
The photon always can be put on shell ($q^2_2 = 0$) at the end. Moreover, for
other choices of $q^2_2$ one can describe processes other than DVCS.
For example, if $q^2_2 > 0$, we can also describe, for example, dilepton
production $\gamma^\ast p \ra \gamma^\ast p \ra l^+l^- p$.

In the next section we show that, in the generalised Bjorken limit, the
amplitude factorises, to leading order in $1/Q^2$, into a convolution
of a perturbative coefficient with a SPDF.


\section{Construction of Compton Amplitude from the Local OPE}
\label{sec_ope}

We now give the proof of factorisation of the leading twist contributions
to the Compton amplitude in terms of the local OPE. A brief outline
is as follows: As with the local
OPE treatment of the total DIS cross-section, where $\xi = t = 0$,
the local OPE gives a simultaneous expansion in inverse powers of
$Q^2$ and inverse powers of $x$. The series thus
converges only in the unphysical region $|x| > 1$ where
the Compton amplitude has no discontinuities (physical cuts). However, one can
use the analytic property of the amplitude to reconstruct the full amplitude
amplitude via a dispersion relation in $x$. The amplitude $T$ can thus be
expressed, to leading order in $1/Q^2$ by the following convolution formula
\be
  T(x, \xi, Q^2, t) \; = \; \int_{-1}^1 \frac{du}{u} \;
   \widehat{C}(\frac{x}{u},\frac{\xi}{u}, Q^2) \; f(u, \xi, t),
\label{factorise}\ee
where $\widehat{C}$
are perturbatively calculable coefficients which can be derived
from the leading Wilson coefficients in the local OPE and where $f$ are the SPDFs.
This factorisation integral is finite for DVCS, $\xi = -x$,
provided that the Wilson coefficients
satisfy a particular condition which we state explicitly below.
This condition is of particular importance since factorisation does not
work if the Wilson coefficients do not satisfy it. When we come to
consider twist-3 contributions in section~\ref{sec_qcd}, we shall need
to define the structure functions in such a way that it is satisfied.

\begin{figure}
\centering{\input{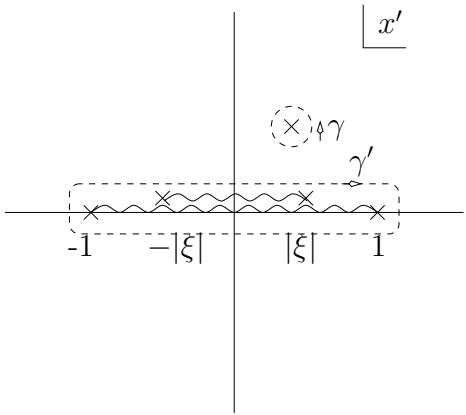}}
\caption{Branch cuts in the complex $\xp$ plane.}
\label{fig_cuts}
\end{figure}

In one of the papers which proved the factorisation formula~(\ref{factorise}),
Radyushkin~\cite{Rad} showed that the coefficient functions are free from
soft singularities. In this context, let us take soft singularities
to be any divergences in the coefficient function $\widehat{C}$ as
$u \ra \pm\xi$, ie. at the endpoints of the ERBL region of the SDPF
(see Appendix) where either parton line in fig.~\ref{fig_fact}
becomes soft. Such soft singularities would take the form of
terms proportional to
$1/(u^2 - \xi^2)$. In the limit $u \ra \pm\xi$, all of these 
should cancel out. We shall show this below explicitly,
starting from the local
OPE. The literature has also been concerned~\cite{CF}
with additional soft singularities
which should appear as the final-state photon goes on shell, $\xi = -x$.
We will show that these are due to terms proportional to $1/(u^2 - x^2)$
which tend towards $1/(u^2 - \xi^2)$ as $x \ra -\xi$.
It was also shown~\cite{CF} that the singularities are integrable and
therefore pose no threat to factorisation, given that the SPDFs are
sufficiently smooth. (See Appendix.)
The local OPE enables us to verify that this is case.

At the end of this section, we consider a simple model for the coefficient
function to show explicitly the cancellation of soft singularities
in the Compton amplitude.
This will enable the reader to follow
the local OPE formalism in a simple example.

\begin{figure}
\centering{\input{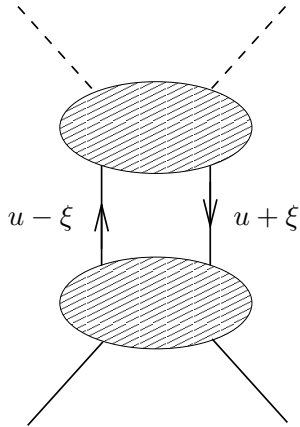}}
\caption{Factorisation of DVCS amplitude in the generalised Bjorken limit.
  The partons' longitudinal momentum fractions are labelled.}
\label{fig_fact}
\end{figure}

To start, we consider the following amplitude in scalar field theory
(fig.~\ref{fig_dvcs}):
\be
  T(q, P, \Delta) \; = \; i \int d^dz \; e^{iq\cdot z} \;
   \mat{p_2}{ T \left( \phi^2(\half z)\, \phi^2(-\half z) \right) }{p_1}.
\label{dvcs_amp}\ee
The amplitude $T(Q^2, t, x, \xi)$ is an analytic
function of its arguments which contains branch cuts in the physical regions.
Since we are going to require later a dispersion relation in the
variable $x$,
we are interested in the branch cuts in the complex $x$ plane which appear
where the following invariant masses are positive:
\bea
  s > 0 & \Rightarrow & 0 < x < 1, \nonumber \\
  u > 0 & \Rightarrow & -1 < x < 0, \nonumber \\
  q^2_1 > 0 & \Rightarrow & ( 0 < x < \xi) \; {\rm OR} \;
   ( \xi < x < 0 ), \\
  q^2_2 > 0 & \Rightarrow & ( 0 < x < -\xi ) \; {\rm OR} \;
   ( -\xi < x < 0 ). \nonumber
\eea
The inequalities are neglecting the corrections of $O(M^2/Q^2)$ and
$O(t/Q^2)$ from eqs.~(\ref{invarts}).
These regions are shown as two wavy lines in the $x$-plane
in fig.~\ref{fig_cuts}.

The dispersion relation is derived by
using Cauchy's theorem in a region away from any
of the abovementioned discontinuities in the $x$ plane (fig.~\ref{fig_cuts}),
we have
\be
  T(x, \xi, Q^2, t) \; = \; \frac{1}{2\pi i} \oint_\gamma \frac{d\xp}{\xp-x}\;
   T(\xp, \xi, Q^2, t).
\ee

Deforming the contour $\gamma$ to $\gamma^\prime$, and under the assumption
that the contour at infinity vanish we have
\be
  T(x, \xi, Q^2, t) \; = \; -\; \frac{1}{\pi} \int_{-1}^1
   \frac{d\xp}{\xp-x}\; \rho_T(\xp, \xi, Q^2, t),
\label{disp_rel}\ee
where
\bea
  2i \; \rho_T(\xp, \xi, Q^2, t) \; = \; - \;
   \raisebox{-0.12in}{$\stackrel{\rm \textstyle disc}{\scriptstyle \xp}$} \;
   T(\xp, \xi, Q^2, t).
\eea
The `discontinuity' is defined as
$\raisebox{-0.08in}{$\stackrel{\rm \textstyle disc}{\scriptstyle z}$}f(z) =
\lim_{\epsilon \ra 0}(f(z + i\epsilon) - f(z - i\epsilon))$. 
One can see from fig.~\ref{fig_cuts} that the discontinuity includes
contributions from s- and u-channels and also from the channels associated
with the invariant masses of the photons, $q^2_{1/2}$.
The latter of these of course vanish in the forward (DIS) limit $\xi \ra 0$.

Since we shall below be obtaining an expansion for $T$ at large $x$ from the
local OPE, we
expand the dispersion relation for large $x$:
\be
  T(x, \xi, Q^2, t) \; = \;
   \frac{1}{\pi} \sum_{j=1}^\infty x^{-j} \; [1 + (-1)^j] \int_0^1
   d\xp \; \xp^{j-1} \; \rho_T(\xp, \xi, Q^2, t)
\label{disp_expand}\ee
Thus, in an expansion at large $|x|$, we can identify the coefficients
with $j$th. Mellin moments of the spectral function $\rho_T$.
So, the steps required in constructing the factorised amplitude are
\begin{enumerate}
 \item Use the local OPE to obtain the large $x$ expansion coefficients and
  identify these with the Mellin moments of the spectral function.
 \item Invert the Mellin moments to obtain the spectral function.
 \item Reconstruct the full, analytic amplitude using the dispersion
  relation~(\ref{disp_rel}).
\end{enumerate}

To implement step one, let us now use the local OPE to obtain the expansion
at large $|x|$ and large $Q^2$, keeping the leading terms in $1/Q^2$.
We do this for the simple case of scalar
fields, to show the general method without any indices.
The time-ordered product of the currents in (\ref{dvcs_amp}) has the following
expansion: 
\bea
  \lefteqn{
   T\left(\phi^2(\half z)\, \phi^2(-\half z)
   \right) \;\; \longrightarrow} & & \label{ofope}\\
  & & \phantom{***}
   -i \; C^{m,j}(z^2) \; z_{\mu_1} \cdots z_{\mu_j} \; \partial^{\mu_1} \cdots
   \partial^{\mu_m} \; :\phi \, \hpart\mbox{}^{\mu_{m+1}} \cdots
   \hpart\mbox{}^{\mu_j} \phi:(0) \; + \; \cdots. \nonumber
\eea
The Wilson coefficients $C^{m,j}(z^2)$ are perturbatively calculable and
will give rise to the `hard' or $Q^2$-dependent part of the factorised
amplitude. The neglected terms are higher order in $z^2$ and are small
as $z^2 \ra 0$. In momentum space, this limit corresponds to $Q^2 \ra \infty$.
Notice also the presence of $m < j$ total derivatives acting on the spin
$j-m$ local composite operators. If we take off-forward matrix elements
of these operators, we can write them in terms of reduced matrix elements
as follows:
\be
  \mat{p_2}{\: :\phi \; i\hpart\mbox{}^{\mu_1} \cdots \,
   i\hpart\mbox{}^{\mu_j} \phi:(0)}{p_1} \; = \; \sum_{{\rm even} \; i=0}^j \;
   \half \Delta^{\mu_1}\, \cdots \, \half\Delta^{\mu_i}\,
   p^{\mu_{i+1}}\cdots p^{\mu_j} \; \langle O^{i,j} \rangle.
\label{ofmat}\ee
By crossing symmetry, only even powers of $\Delta$ are included in this
decomposition.
Taking the Fourier transform of the off-forward matrix element of the 
expansion (\ref{ofope}) and using (\ref{ofmat}) we get the
amplitude in momentum space
\be
  T(x, \xi, Q^2, t) \; \stackrel{x, Q^2 \ra \infty}{\longrightarrow} \;
   \sum_{{\rm even} \;j=2}^\infty x^{-j}
   \sum_{{\rm even} \; m=0}^j \widetilde{C}^{m,j}(Q^2) \; \xi^m \;
   \sum_{{\rm even} \; i=0}^{j-m} \xi^{i} \langle O^{i, j-m} \rangle ,
\label{ofope2}\ee
where
\be
  \widetilde{C}^{m,j}(Q^2) \; = \; \left( Q^2 \frac{d}{dQ^2} \right)^j
   \int d^dz \; e^{iq\cdot z} \; C^{m,j}(z^2).
\ee
Comparing eqs.~(\ref{ofope2}) and (\ref{disp_expand}) we can read off 
coefficients of $x^{-j}$ and we get
\be
  \frac{1}{\pi} \int_0^1 dx \; x^{j-1} \; \rho_T(x, \xi, Q^2, t) \; = \;
   \frac{1}{2}\sum_{{\rm even} \; m=0}^j \widetilde{C}^{m,j}(Q^2) \; \xi^m \;
   \sum_{{\rm even} \; i=0}^{j-m} \xi^{i} \langle O^{i,j-m} \rangle.
\label{fact_mellin}\ee
Eq.~(\ref{fact_mellin}) expresses the factorisation of the moments
of the spectral function. The hard, perturbative physics is encoded in
the Wilson coefficients, $\widetilde{C}$, while the non-perturbative physics
is in the reduced matrix elements. Notice that the reduced matrix elements
appear in polynomials of $\xi$ of order $j-m$. In the appendix on SPDFs
it is shown that these polynomials are equal to moments of the SPDFs:
\be
  f^j(\xi, t) \; = \;
   \int_0^1 du \; u^{j-1} \; f(u, \xi, t) \; = \; \frac{1}{2}
   \sum_{{\rm even}\; i=0}^j \xi^i \; \langle O^{i,j} \rangle.
\label{def_skew}\ee
With this identification, (\ref{fact_mellin}) becomes
\be
  \frac{1}{\pi} \int_0^1 dx \; x^{j-1} \; \rho_T(x, \xi, Q^2, t) \; = \;
   \sum_{{\rm even} \; m=0}^j \widetilde{C}^{m,j}(Q^2) \; \xi^m \;
   f^{j-m}(\xi, t).
\label{fact_mellin2}\ee

For step two, we need to invert the Mellin transform in
eq.~(\ref{fact_mellin2}). 
One can verify that this is solved by the following expression:
\bea
  \frac{1}{\pi} \; \rho_T(x, \xi, Q^2, t) & = &
   \theta(x > 0) \int_x^1 \frac{du}{u} \;
   C(\frac{x}{u},\frac{\xi}{u}, Q^2) \; f(u, \xi, t) \nonumber \\
  &-&\theta(x < 0) \int_{-1}^x \frac{du}{u} \;
   C(\frac{x}{u},\frac{\xi}{u}, Q^2)
   \; f(u, \xi, t) \nonumber \\
  & - & \frac{1}{2}\;\theta(0 < x < |\xi|) \int_{-1}^1 \frac{du}{u} \;
   D(\frac{x}{u},\frac{\xi}{u}, Q^2) \; f(u, \xi, t) \nonumber \\
  & + & \frac{1}{2} \; \theta(-|\xi| < x < 0) \int_{-1}^1 \frac{du}{u} \;
   D(\frac{x}{u},-\,\frac{\xi}{u}, Q^2) \; f(u, \xi, t),
\label{geko}\eea
where the coefficient functions $C$ and $D$ have been introduced so that
\bea
  \int_0^1 x^{j-1} \;C(x, \xi, Q^2)
   & = & \sum_{{\rm even}\; m=0}^\infty \widetilde{C}^{m,j}(Q^2)
   \; \xi^j, \label{series_one}\\
  \int_0^1 x^{j-1} \;D(x\xi, \xi, Q^2)
   & = & \sum_{{\rm even}\; m=2}^\infty \widetilde{C}^{m+j, j}(Q^2)
   \; \xi^j\label{series_two}.
\eea
Notice that we have four terms in the spectral function. These correspond
to the s-channel, u-channel and $q^2_{1/2}$-channel cuts respectively.

Finally, for step three, we can insert~(\ref{geko}) into the dispersion
relation~(\ref{disp_rel}) to obtain our final form for the factorised
amplitude:
\be
  T(Q^2, t, x, \xi) \; = \; \int_{-1}^1 \frac{du}{u} \;
   \widehat{C}(Q^2, \frac{x}{u}, \frac{\xi}{u}) \; f(u, \xi, t),
\label{fact_amp}\ee
where
\bea
  \widehat{C}(Q^2, \frac{x}{u}, \frac{\xi}{u}) & = & -
   \; \frac{1}{2} \int_0^u d\xp
   \left[\frac{1}{\xp - x} \; + \; \frac{1}{\xp + x}\right]
   C(\frac{\xp}{u},\frac{\xi}{u}, Q^2) \nonumber \\
  & + & \frac{1}{2} \int_0^\xi d\xp
   \left[\frac{1}{\xp - x} \; + \; \frac{1}{\xp + x}\right]\;
   D(\frac{\xp}{u}, \frac{\xi}{u}, Q^2).
\label{final_form}\eea

Three points are now in order: First, the soft singularities can be
shown to cancel out for $x \neq \pm\xi$:
The key point about the infinite series
in~(\ref{series_one})-(\ref{series_two}) is that they diverge as
$|\xi| \ra 1$ so that they each behave as $1/(1-\xi^2)$. In the convolution
formula in eq.~(\ref{geko}), these singularities appear as $u \ra \pm\xi$
so that, typically, 
\be
  C(\frac{x}{u},\frac{\xi}{u}, Q^2) \; \sim \;
   D(\frac{x}{u},\frac{\xi}{u}, Q^2) \; \sim \;
   \frac{1}{u^2 - \xi^2}.
\ee
Since the ERBL region of the SPDFs is $-|\xi| < u < |\xi|$, $C$ and $D$
diverge at the endpoint of the ERBL region. However, in this case the
singularities cancel between the functions $C$ and $D$. We can check
this by taking the limit $u \ra \xi$ of the coefficient
function $\widehat{C}$ to get
\bea
  \lim_{u\ra\xi} \; \widehat{C}(\frac{x}{u}, \frac{\xi}{u}, Q^2) \; = \;
   -\int_0^\xi d\xp \left[\frac{1}{\xp - x} \; + \; \frac{1}{\xp + x}\right]\;
   \lim_{u\ra\xi}\left( C(\frac{\xp}{\xi}, \frac{\xi}{u}, Q^2) \; - \;
    D(\frac{\xp}{\xi}, \frac{\xi}{u}, Q^2)\right).
\eea
$C$ and $D$ both have the same divergent behaviour, as can be deduced
from the infinite series (\ref{series_one})-(\ref{series_two}), which cancels
between them.

The second point concerns the presence of terms that go like $1/(u^2 - x^2)$
which arise from the first integral in eq.~(\ref{final_form}). For example,
if $C(x, \xi) = \delta(1-x)$, we would have
\be
  \widehat{C}(\frac{x}{u}, \frac{\xi}{u}) \; = \; - \; \frac{u^2}{u^2 - x^2}.
\label{sing}\ee
These are the additional soft singularities which, as mentioned before,
arise as either photon goes on shell, $\xi \ra \pm x$.
It is quite easy to show that we will not get terms any more singular than
in eq.~(\ref{sing}).
As mentioned above, in the DVCS limit $x \ra -\xi$, this kind
of soft singularity approaches the endpoints of the ERBL region $u = \pm\xi$.
However, given a smooth enough SPDF (see appendix), the singularity 
remains integrable. We could not, however, tolerate terms more singular than
in eq.~(\ref{sing}).

The third point is that if we put one of the photons on shell
by taking the limit $\xi \ra \pm x$ in the factorised
amplitude~(\ref{fact_amp}), it appears that
the coefficient function might diverge logarithmically.
This happens when either limit of
the second integral over $D$ in eq.~(\ref{final_form})
is trapped at either of the two poles in
$x \pm \xp$ and indicates the presence of $\ln^p(1 \pm \xi/x)$ terms.
However, no such difficulty occurs provided that $D(x, \xi, Q^2) \ra 0$
as $\xi \ra \pm x$. A sufficient condition for this to happen can be
derived in terms of the Mellin moment of $D$ in eq.~(\ref{series_two}):
As $j$ becomes larger, the integral samples the region $\xi = \pm x$ with
a higher weighting. If $D(x, \xi, Q^2)$ goes like $\xi - x$ in this
region, one can show that
\bea
  \int_0^1 dx \; x^{j-1} \; D(x\xi, \xi, Q^2) &
   \sim & \int_0^1 dx \; x^{j-1} \; (1 - x) \nonumber \\
  & = & \frac{1}{j(j+1)} \; \stackrel{j \ra \infty}{\lngra} \; \frac{1}{j^2}.
\eea
Thus, from the r.h.s. of eq.~(\ref{series_two}), a sufficient condition
for the absence of $\ln^p(1 \pm \xi/x)$ terms is
\be
  \widetilde{C}^{m+j,j} \; \stackrel{j \ra \infty}{\lngra} \; \frac{1}{j^2}.
\label{condition}\ee

Eq.~(\ref{condition}) is crucially important to factorisation as treated
in the local OPE: If it is satisfied then factorisation is valid in the
DVCS kinematics.
All NLO calculations~\cite{OJ,Man,BM} of twist-2 contributions that have been
presented are of course consistent with~(\ref{condition}).

\subsection*{A simple model}

To illustrate the application of the local OPE, let us now consider a simple
example which has all the features of a factorisable amplitude. We choose
a simple but non-trivial model for the Wilson coefficients which satisfy
eq.~(\ref{condition}). Then we perform steps two and three in the method
described above to obtain the factorised form for the `amplitude'.

Suppose, then, that we performed an OPE and obtained the following expansion
\be
  T(x, \xi) \; \stackrel{Q^2, x \ra \infty}{\lngra} \;
   \sum_{{\rm even}\; j = 2}^\infty x^{-j} \sum_{{\rm even}\; m = 0}^j
   \xi^m \; \widetilde{C}^{m,j}
   \; f^{j-m}(\xi),
\ee
where 
\be
  \widetilde{C}^{m,j} \; = \; \frac{1}{(j+1)(j+2)},
\ee
and $f^j(\xi)$ is the $j$th. moment of the skewed PDF.

For step two, we take the coefficients of $x^{-j}$ and perform the inverse
Mellin transform w.r.t. $j$ to obtain:
\bea
  \frac{1}{\pi} \; \rho_T(x, \xi) & = & \theta(x > 0) \; \int_x^1
   \frac{du}{u} \; C(\frac{x}{u}, \frac{\xi}{u}) \; f(u, \xi) \nonumber \\
  & - & \theta(\xi > x > 0) \; \int_0^1 \frac{du}{u} \;
   D(\frac{x}{u}, \frac{\xi}{u}) \; f(u, \xi) \; + \; \cdots \; ,
\eea
where the neglected terms are those which are required to make $\rho_T$ an
odd function of $x$. We also take $\xi > 0$. The coefficient functions
are given by\footnote{%
The inverse Mellin transform of $1/((j+1)(j+2))$ is the function $u(1-u)$.}
\bea
  C(x, \xi) & = & \sum_{{\rm even} \; m = 0}^\infty \xi^m \; x(1-x)
   \nonumber \\
  & = & u(1-u) \; \frac{1}{1 - \xi^2}, \\
  D(x, \xi) & = & \sum_{{\rm even} \; m = 2}^\infty \xi^m \;
   \frac{x}{\xi}\left(1 - \frac{x}{\xi}\right) \nonumber \\
  & = & \frac{x}{\xi}\left(1 - \frac{x}{\xi}\right) \;
   \frac{\xi^2}{1 - \xi^2}.
\eea

For step three, we replace $\rho_T$ into the dispersion relation. This
gives us a double integral. Performing the integral over $\xp$ we arrive
at the final convolution integral:
\bea
  T(x, \xi) & = & - \frac{1}{2} \int_{-1}^1 \frac{du}{u} \left[
   \left( 1 \; + \; \frac{x}{u} \; \ln\left(\frac{1 - u/x}{1 + u/x}\right)
   \; - \; \frac{x^2}{u^2} \; \ln\left(1 - \frac{u^2}{x^2}\right) \right)
   \frac{u^2}{u^2 - \xi^2} \right. \nonumber \\
  & & \phantom{- \frac{1}{2} \int_{-1}^1 \frac{du}{u}} \left. - \;
   \left( 1 \; + \; \frac{x}{\xi} \; \ln\left(\frac{1 - \xi/x}{1 + \xi/x}\right)
   \; - \; \frac{x^2}{\xi^2} \; \ln\left(1 - \frac{\xi^2}{x^2}\right) \right)
   \frac{\xi^2}{u^2 - \xi^2} \right] f(u, \xi). \nonumber \\
\eea
The coefficient of $f$ in the integral can be seen to have the property
that it is free of soft singularities, ie. as $u \ra \pm\xi$. It is also
finite in the DVCS kinematics $\xi = \pm x$.


\section{The Local OPE in QCD}
\label{sec_qcd}

In the previous section, we applied the local OPE to the analogue
of DVCS in scalar field theory. We now consider the QCD case. First, it
will be important to make some comments about how to organise the local OPE.
It is important to keep conservation of both electromagnetic
currents explicit. To this end, we shall start the section by writing
down a general decomposition
of the Compton amplitude in terms of Lorentz scalar structure functions,
keeping the photons of unequal mass. For simplicity we
shall retrict our attention to the case of a scalar target which,
for practical
purposes, can be an unpolarised proton. We also work in the approximation
of neglecting any terms suppressed by $O(t/Q^2)$. All structure functions
are such that electromagnetic current conservation is obeyed for each.

Afterwards,
we shall derive the local OPE for the product of electromagnetic
currents at the tree-level
approximation. We show explicitly that current conservation is satisfied
because of a cancellation between parity-even and -odd, twist-3 operators.
This has been emphasised in ref.~\cite{BM2}. The first purpose of this section
is to review briefly their work. Another important observation of
ref.~\cite{BM2} was that the operator identities which allowed the twist-3
terms to cancel also enable us to relate the twist-3 operators
to purely twist-2 terms.
Once the OPE is recast in terms of these twist-2 pieces, it can be seen
that even and odd parity sectors satisfy current conservation separately.
This means that the OPE does indeed enable
us to partition the operators of opposite parity from each other
in different Lorentz structures.

Once the OPE has been treated, we will then take off-forward
matrix elements of the OPE expansion in unpolarised proton states. This
will enable us to give tree-level predictions for the structure functions
in terms of integral convolutions of twist-2 SPDFs with coefficient functions.
This has also been achieved in the literature before~\cite{BM2,BMKS,Tw3}.
However, the structure
functions which receive contributions only from twist-3 pieces were found
to diverge logarithmically in the DVCS limit. In our treatment, we give
a different definition of the structure functions for which the tree-level
convolution integrals are finite.

\subsection*{Structure functions}

Let us write down the decomposition of the Compton amplitude tensor
$T^{\mu\nu}$ for the case of a scalar target. We only include those Lorentz
structures which are of even parity and satisfy current conservation, ie.
\be
  (q + \half \Delta)_\mu T^{\mu\nu}(q, p, \Delta) \; = \;
  (q - \half \Delta)_\nu T^{\mu\nu}(q, p, \Delta) \; = \; 0.
\ee
It will be convenient to introduce the following four-vectors:
\bea
  \hat{q}^\mu & = & q^\mu \; + \; x \, p^\mu, \\
  \Delta_\perp^\mu & = & \Delta^\mu \; - \; 2\xi \, p^\mu.
\eea
These are such that $q\cdot \Delta_\perp = 0$ while
$\hat{q}^2$ \& $\Delta_\perp^2 = O(t)$.
Thus, keeping only terms of order $\Delta_\perp$ and neglecting orders
$\Delta_\perp^2$ and $t$, the most general structure function decomposition
is as follows:\footnote{%
We use the convention $a^{\{\mu} b^{\nu\}} = a^\mu b^\nu + a^\nu b^\mu$ and
$a^{[\mu} b^{\nu]} = a^\mu b^\nu - a^\nu b^\mu$.}
\bea
  T^{\mu\nu}(q, p, \Delta) & = &
   \left( -g^{\mu\nu} \; + \; \frac{1}{q\cdot p} \, \left(
   p^{\{\mu} \, q^{\nu\}} \; + \; 2x \, p^\mu \, p^\nu \; + \;
   \half \, p^{[\mu} \, \Delta_\perp^{\nu]} \right) \right)
   T_T(x, \xi, Q^2, t) \nonumber \\
  & + & \frac{1}{4q\cdot p}
   \left( p^\mu \; + \; \frac{\hat{q}^\mu + \half \Delta_\perp^\mu}{x - \xi}
   \right)
   \left( p^\nu \; + \; \frac{\hat{q}^\nu - \half \Delta_\perp^\nu}{x + \xi}
   \right) T_L(x, \xi, Q^2, t) \nonumber \\
  & + & \frac{1}{2\xi\, q\cdot p}
   \left[ \Delta_\perp^\mu
   \left( p^\nu \; + \; \frac{\hat{q}^\nu}{x + \xi} \right) \; + \;
   \Delta_\perp^\nu \left( p^\mu \; + \; \frac{\hat{q}^\mu}{x - \xi} \right)
   \right] T^-_3(x, \xi, Q^2, t) \nonumber \\
  & + & \frac{1}{2 q\cdot p}
   \left[ \Delta_\perp^\mu
   \left( p^\nu \; + \; \frac{\hat{q}^\nu}{x + \xi} \right) \; - \;
   \Delta_\perp^\nu \left( p^\mu \; + \; \frac{\hat{q}^\mu}{x - \xi} \right)
   \right] T^+_3(x, \xi, Q^2, t) \nonumber \\
  & + & O(\Delta^2_\perp, t).
\label{lor_struct}\eea
The structure functions $T_L$ and $T^+_3$ are both even under crossing symmetry
and are therefore even functions of $x$ and $\xi$. $T_T$ and
$T^-_3$ are odd under crossing symmetry and are odd functions of $x$ and
even functions of $\xi$. 
In the forward (DIS) limit, $\xi$ \& $\Delta_\perp \ra 0$, the above
decomposition reduces to just two terms with $T_T$ and $T_L$, the familiar
transverse and longitudinal structure functions.

The reader may object however that, when we look at the case of DVCS,
$\xi = -x$ or equivalently $\xi = x$, the Lorentz structures multiplying
$T_L$ and $T^\pm_3$ become large.
Other authors have not used these definitions
for this reason. However, the divergence is physical and should not be
included in the definition of the structure functions. As a result,
the factorisation formulae previously found~\cite{BM2} for the alternative
definitions of $T^\pm_3$
have been found to be divergent. The divergence does not appear in physical
amplitudes because, at this order in $\Delta_\perp$, the physical polarisation
vectors of the final-state real photon are perpendicular to the vector
\be
  p^\nu \; + \; \frac{\hat{q}^\nu - \half \Delta_\perp^\nu}{x + \xi}.
\ee

\subsection*{Local OPE at Tree Level}

In order to obtain predictions for the four structure functions in terms
of skewed PDFs to leading order in the strong coupling, we now use the OPE
method introduced in the last section to demonstrate factorisation.
What we shall find is that $T_L$ vanishes while $T_T$ is a simple convolution
over the twist-2 unpolarised
SPDF. $T^\pm_3$ are more involved, and arise from twist-3 SDPFs, but which
themselves can be expressed as 
convolutions of coefficient functions with the twist-2 SPDF.

To begin, we write down the tree-level
OPE for the product of currents explicitly, to leading twist and for
one flavour of quark:
\bea
  \lefteqn{T(j^\mu(y)\,j^\nu(z))} & & \nonumber \\
  & & \simeq \;\; \frac{i}{2\pi^2} \; \frac{(y-z)_\alpha}{(y-z)^4} \;
   \left[ \psibar(y)\, \gamma^\mu \gamma^\alpha \gamma^\nu \, \psi(z) \; -
   \; \psibar(z)\, \gamma^\nu \gamma^\alpha \gamma^\mu \, \psi(y) \right]
   \nonumber \\
  & & = \;\; \frac{i}{2\pi^2} \; \frac{(y-z)_\alpha}{(y-z)^4} \nonumber \\
  & & \phantom{**} \times \left[ S^{\mu\nu\alpha\beta}
   \sum_{{\rm even} \; j = 2}^\infty \frac{1}{(j-1)!} \;
   (z-y)_{\mu_1} \cdots (z-y)_{\mu_{j-1}} \;\psibar \, \gamma_\beta
   \hpart\mbox{}^{\mu_1} \cdots \hpart\mbox{}^{\mu_{j-1}} \, \psi(\half(y+z))
   \right. \nonumber \\
  & & \phantom{**}
   \left. - \; i\epsilon^{\mu\nu\alpha\beta} \sum_{{\rm odd} \; j=1}^\infty
   \frac{1}{(j-1)!} \; (z-y)_{\mu_1} \cdots (z-y)_{\mu_{j-1}} \;
   \psibar \, \gamma_\beta \gamma_5
   \hpart\mbox{}^{\mu_1} \cdots \hpart\mbox{}^{\mu_{j-1}} \, \psi(\half(y+z))
   \right], \nonumber \\
\label{tree_ope}\eea
where
\be
  S^{\mu\nu\alpha\beta} \; = \; g^{\mu\alpha}\;g^{\nu\beta} \; + \;
   g^{\mu\beta}\;g^{\nu\alpha} \; - \; g^{\mu\nu}\;g^{\alpha\beta}.
\ee
Despite the expansion into local operators in the second step, current
conservation is still satisfied up to equation of motion pieces. Explicitly,
\bea
  \lefteqn{\partial_\mu T(j^\mu(y)\,j^\nu(z)) \; \simeq} & & \nonumber \\
  & &  \;\; \frac{i}{2\pi^2} \; \frac{(y-z)_\alpha}{(y-z)^4} \;
   \left[ \sum_{{\rm odd} \; j = 1}^\infty
   \frac{1}{(j-1)!} \; (z-y)_{\mu_1} \cdots (z-y)_{\mu_{j-1}}
   \right. \nonumber \\
  & & \phantom{ \;\; \frac{i}{2\pi^2} \; \frac{(y-z)_\alpha}{(y-z)^4} \;}
   \times \; \left( \psibar \, \gamma^{[\nu} \hpart\mbox{}^{\alpha]}
   \hpart\mbox{}^{\mu_1} \cdots \hpart\mbox{}^{\mu_{j-1}} \, \psi(\half(y+z)) \right.
   \nonumber \\
  & & \phantom{ \;\; \frac{i}{2\pi^2} \; \frac{(y-z)_\alpha}{(y-z)^4} \;}
   \left. - \;  \frac{i}{2} \epsilon^{\nu\alpha\mu\beta} \;
   \partial_\mu \; \psibar \, \gamma_\beta \gamma^5
   \hpart\mbox{}^{\mu_1} \cdots \hpart\mbox{}^{\mu_{j-1}} \, \psi(\half(y+z)) \right)
   \nonumber \\
  & & \phantom{ \;\; \frac{i}{2\pi^2} \; \frac{(y-z)_\alpha}{(y-z)^4} \;}
   + \sum_{{\rm even} \; j=2}^\infty
   \frac{1}{(j-1)!} \; (z-y)_{\mu_1} \cdots (z-y)_{\mu_{j-1}} \nonumber \\
  & & \phantom{ \;\; \frac{i}{2\pi^2} \; \frac{(y-z)_\alpha}{(y-z)^4} \;}
   \times \; \left( \frac{1}{2} \; \partial^{[\alpha} \; \psibar
   \, \gamma^{\nu]}
   \hpart\mbox{}^{\mu_1} \cdots \hpart\mbox{}^{\mu_{j-1}} \, \psi(\half(y+z)) \right.
   \nonumber \\
  & & \phantom{ \;\; \frac{i}{2\pi^2} \; \frac{(y-z)_\alpha}{(y-z)^4} \;}
   \left. \left. - \;  i \epsilon^{\nu\alpha\mu\beta} \;
   \psibar \, \gamma_\beta \gamma^5 \hpart\mbox{}\!\!_\mu
   \hpart\mbox{}^{\mu_1} \cdots \hpart\mbox{}^{\mu_{j-1}}
   \, \psi(\half(y+z)) \right) \right]\\
  & & \;\; = 0. \nonumber
\eea
The expression vanishes because the pairs of terms in parentheses cancel,
due to the following identities, which are satisfied up to an equation of
motion, and neglecting quark masses and the strong coupling:
\bea
  \psibar \, \gamma^{[\mu} \hpart\mbox{}^{\nu]}
   \hpart\mbox{}^{\mu_1} \cdots \hpart\mbox{}^{\mu_{j-1}} \, \psi & = &
   \frac{i}{2} \epsilon^{\mu\nu\alpha\beta} \;
   \partial_\alpha \; \psibar \, \gamma_\beta \gamma^5
   \hpart\mbox{}^{\mu_1} \cdots \hpart\mbox{}^{\mu_{j-1}} \, \psi,
   \label{tw_3_id1}\\
  \frac{1}{2} \; \partial^{[\mu} \; \psibar
   \, \gamma^{\nu]}
   \hpart\mbox{}^{\mu_1} \cdots \hpart\mbox{}^{\mu_{j-1}} \, \psi & = &
   i \epsilon^{\mu\nu\alpha\beta} \;
   \psibar \, \gamma_\alpha \gamma^5 \hpart\mbox{}_\beta
   \hpart\mbox{}^{\mu_1} \cdots \hpart\mbox{}^{\mu_{j-1}} \, \psi.
   \label{tw_3_id2}
\eea

At first sight, it seems that it will not be possible to arrange for
the quark operators of each parity to appear in two separate Lorentz
structures, each satisfying current conservation. This is not the case,
however~\cite{BM2}. The quark operators, as they appear in the OPE
in~(\ref{tree_ope}) can be decomposed in terms of purely twist-2
(totally symmetric) operators, in a series with increasing numbers of total
derivatives. To show this, we start by taking the quark operators and
decomposing them into two pieces, of twist-2 and twist-3
(one pair of indices antisymmetrised) terms as follows:
\be
  O_i^{(\alpha)\mu_1 \cdots \mu_{j-1}} \;=\; O_i^{\alpha\mu_1 \cdots
   \mu_{j-1}} \; + \; \frac{2(j-1)}{j} \;
   \raisebox{-0.12in}{$\stackrel{\textstyle \rm S}{\scriptstyle \alpha \{\mu\}}
   $} \;
   \raisebox{-0.085in}{$\stackrel{\textstyle \rm A}{\scriptstyle \alpha \mu_1}
   $} \; O_i^{(\alpha)\mu_1 \cdots \mu_{j-1}},
\label{twisty}\ee
where the S and A operators symmetrise and antisymmetrise,
respectively, the indices listed in subscript. Also, $O_i \in \{O, O_5\}$ and
\bea
  O^{(\alpha)\mu_1 \cdots \mu_{j-1}} & = &
   \psibar \, \gamma^\alpha
   \hpart\mbox{}^{\mu_1} \cdots \hpart\mbox{}^{\mu_{j-1}} \, \psi, \\
  O_5^{(\alpha)\mu_1 \cdots \mu_{j-1}} & = &
   \psibar \, \gamma^\alpha \, \gamma_5
   \hpart\mbox{}^{\mu_1} \cdots \hpart\mbox{}^{\mu_{j-1}} \, \psi, \\
  O_i^{\alpha \mu_1 \cdots \mu_{j-1}} & = &
   \raisebox{-0.12in}{$\stackrel{\textstyle \rm S}{\scriptstyle \alpha \{\mu\}}
   $} \; O_i^{(\alpha) \mu_1 \cdots \mu_{j-1}}.
\eea
In order to express~(\ref{twisty}) purely in terms of twist-2 pieces,
one performs this iterative procedure: Substitute the twist-3 terms
in~(\ref{twisty}) with eqns.~(\ref{tw_3_id1})-(\ref{tw_3_id2}); decompose
the substituted terms again using~(\ref{twisty}); repeat the first two steps
until there are no more twist-3 terms. The resulting series can be
summed to give
\bea
  O^{(\alpha)\mu_1 \cdots \mu_{j-1}} & = & O^{\alpha\mu_1 \cdots \mu_{j-1}}
   \nonumber\\
  & + & \frac{1}{j} \;
   \raisebox{-0.12in}{$\stackrel{\textstyle \rm S}{\scriptstyle \alpha \{\mu\}}
   $} \left[ \sum_{{\rm even} \; k=0}^{j-2} \left(\frac{i}{2}\right)^{k+1}
   (j-k-1) \; \partial^{\mu_1} \cdots \partial^{\mu_k} \,
   \epsilon^{\alpha\mu_{k+1}\beta}_{\phantom{\alpha\mu_{k+1}\beta}\gamma} \,
   \partial_\beta \; O_5^{\gamma\mu_{k+2} \cdots \mu_{j-1}}\right. \nonumber \\
  & + & \sum_{{\rm odd}\; k=1}^{j-3} \left(\frac{i}{2}\right)^{k+1}
   (j-k-1) \; \partial^{\mu_1} \cdots \partial^{\mu_{k-1}} \nonumber \\
  & & \left. \phantom{**\frac{!}{!}} \times \; \{
   \partial^{\mu_k}\,\partial^{\mu_{k+1}} \; O^{\alpha\mu_{k+2}\cdots\mu_{j-1}}
   \; - \; (-\Box\,g^{\alpha\mu_k} \; + \; \partial^\alpha\,\partial^{\mu_k})\;
   O^{\mu_{k+1}\cdots\mu_{j-1}} \} \right], \nonumber \\
\label{greg}\eea
and the same result again with $O$ and $O_5$ swapped.
All derivatives in this expression are total derivatives acting on the
local operators. The expression differs slightly from that in ref.~\cite{BM2}
which does not contain the term with the $\Box$, but this
term can be discarded anyway since
\be
  \Box \psibar \; \cdots \; \psi \; = \; -4 \; \psibar \hpart\mbox{}^\lambda
   \hpart\mbox{}_\lambda \cdots \; \psi \; + \; {\rm EOM}.
\ee
Thus such terms correspond to traces and are therefore higher twist.

By writing the matrix elements
of the twist-2 operators in terms of reduced matrix elements, one can
proceed as in the previous section and derive the form of the OPE in momentum
space. The matrix elements in unpolarised proton states are
\bea
  \mat{p_2}{O^{\mu_1 \cdots \mu_j}}{p_1} & = & (i)^{j-1}
   \raisebox{-0.12in}{$\stackrel{\textstyle \rm S}{\scriptstyle \{\mu\}}
   $} \sum_{{\rm even}\;k = 0}^j \half \Delta^{\mu_1} \; \cdots \;
   \half \Delta^{\mu_{k}} \; p^{\mu_{k+1}} \; \cdots p^{\mu_j} \; \langle
   O^{m,j}\rangle, \nonumber \\
  \mat{p_2}{O_5^{\mu_1\cdots\mu_j}}{p_1} & = & 0.
\label{mat_elem}\eea
The matrix elements of the twist-2 operator $O_5$ vanish because it
is not possible to write down a completely symmetric, odd-parity Lorentz
structure built only out of the vectors $p$ and $\Delta$. Now we can find the
form of the matrix elements of the quark operators including twist-3
contributions. It will
be convenient for our purposes to write these not in terms of $\Delta$ but
of $\Delta_\perp$, and keep only those terms up to $O(\Delta_\perp)$.
Using eqs.~(\ref{greg}) and~(\ref{mat_elem}), we get
\bea
  \mat{p_2}{O^{(\alpha)\mu_1 \cdots \mu_{j-1}}}{p_1} & = &
   i^{j-1} \; p^\alpha \, p^{\mu_1} \, \cdots \, p^{\mu_{j-1}}\; O^L_j(\xi)
   \nonumber \\
  & + & i^{j-1} \; p^\alpha \,
   \raisebox{-0.12in}{$\stackrel{\textstyle \rm S}{\scriptstyle \{\mu\}}$}
   \, \frac{\Delta_\perp^{\mu_1}}{2\xi} \, p^{\mu_2} \,\cdots\, p^{\mu_{j-1}}
   \; O^{L2}_j (\xi) \nonumber \\
  & + & i^{j-1} \; \frac{\Delta_\perp^{\alpha}}{2\xi} \, p^{\mu_1} \, \cdots \,
   p^{\mu_{j-1}} \; O^T_j(\xi), \label{poop1}\\
  \mat{p_2}{O_5^{(\alpha)\mu_1 \cdots \mu_{j-1}}}{p_1} & = &
   \half i^{j-2} \;
   \raisebox{-0.12in}{$\stackrel{\textstyle \rm S}{\scriptstyle \{\mu\}}$}
   \epsilon^{\alpha\mu_1 \beta\gamma} \, \Delta^\perp_\beta
   p_\gamma \, p^{\mu_2} \, \cdots \, p^{\mu_{j-1}} \;
   \widetilde{O}^T_j(\xi), \label{poop2}
\eea
where
\bea
  O^L_j(\xi) & = & \sum_{{\rm even}\;k = 0}^j \xi^k \;
   \langle O^{m,j}\rangle, \\
  O^T_j(\xi) & = & \frac{1}{j} \; \xi\frac{d}{d\xi} \; O^L_j(\xi) \; - \;
   \left( 1 \; - \; \frac{1}{j} \; \xi\frac{d}{d\xi} \right)
   \sum_{{\rm even}\;k = 2}^j \xi^k \; O^L_{j-k}(\xi), \\
  O^{L2}_j(\xi) & = & \frac{j-1}{j} \; \xi\frac{d}{d\xi} \; O^L_j(\xi)
   \; + \; \left( 1 \; - \; \frac{1}{j} \; \xi\frac{d}{d\xi} \right)
   \sum_{{\rm even}\;k = 2}^j \xi^k \; O^L_{j-k}(\xi), \\
  \widetilde{O}^T_j(\xi) & = & \frac{1}{j}
   \left( j - 1 - \xi\frac{d}{d\xi} \right)
   \sum_{{\rm even}\;k = 0}^{j-1} \xi^k \; O^L_{j-k-1}(\xi).
\eea

All that remains for us to obtain the tree-level, large-$x$ expansion is
to take the Fourier transform of the off-forward matrix elements of
the r.h.s. of eq.~(\ref{tree_ope}), then substitute the matrix elements
in eqs.~(\ref{poop1}) and~(\ref{poop2}). This we write below, organised so that
we can read off the structure functions easily:
\bea
  \lefteqn{i \int d^4z \; e^{iq\cdot z} \;
   \mat{p_2}{T( \, j^\mu(\half z) \, j^\nu(-\half z) \, )}{p_1}} \nonumber \\
  & = &
   \left( -g^{\mu\nu} \; + \; \frac{1}{q\cdot p} \, \left(
   p^{\{\mu} \, q^{\nu\}} \; + \; 2x \, p^\mu \, p^\nu \; + \;
   \half \, p^{[\mu} \, \Delta_\perp^{\nu]} \right) \right)
   \sum_{{\rm even}\; j = 2}^\infty x^{-j} \;
   O^L_j(\xi) \nonumber \\
  & + &
   \frac{1}{2\xi\, q\cdot p} \left[ \Delta_\perp^\mu
   \left( p^\nu \; + \; \frac{\hat{q}^\nu}{x + \xi} \right) \; + \;
   \Delta_\perp^\nu \left( p^\mu \; + \; \frac{\hat{q}^\mu}{x - \xi} \right)
   \right]  \sum_{{\rm odd}\; j = 1}^\infty x^{-j} \;
   ( O^T_{j+1}(\xi) \; + \; \xi^2 \, \widetilde{O}^T_j(\xi) )
   \nonumber \\
  & + & \frac{1}{2q\cdot p}
   \left[ \Delta_\perp^\mu
   \left( p^\nu \; + \; \frac{\hat{q}^\nu}{x + \xi} \right) \; - \;
   \Delta_\perp^\nu \left( p^\mu \; + \; \frac{\hat{q}^\mu}{x - \xi} \right)
   \right] \sum_{{\rm even}\; j = 2}^\infty x^{-j} \;
   ( O^T_j(\xi) \; + \; \widetilde{O}^T_{j+1}(\xi) ). \nonumber \\
\eea
Note that the structure corresponding to $T_T$ receives contributions
only from the twist-2 pieces of the matrix elements $O^L$. Also,
$T_L$ vanishes as claimed, while $T^\pm_3$ receive twist-3 contributions
only. These we write out explicitly so that we can check that they
obey the factorisation condition of eq.~(\ref{condition}) for DVCS:
\bea
  \lefteqn{O^T_{j+1}(\xi) \; + \; \xi^2 \, \widetilde{O}^T_j(\xi)}
   \nonumber \\ 
  & \phantom{****} = &
   \left(\frac{1}{j} \; - \; \frac{1}{j(j+1)}\; \xi\frac{d}{d\xi} \right)
   \sum_{{\rm even}\; k = 2}^{j+1} \xi^k \; O^L_{j-k+1}(\xi) \; + \;
   \frac{1}{j+1} \; \xi\frac{d}{d\xi} O^L_{j+1}(\xi) \nonumber \\
  & \phantom{****} = & \sum_{{\rm even}\; k = 0}^{j+1} \xi^k \;
   \left\{ \widetilde{C}^{k, j}_{a-} \; O^L_{j-k+1}(\xi) \; + \;
   \widetilde{C}^{k,j}_{b-} \; \xi\frac{d}{d\xi} \; O^L_{j-k+1}(\xi) \right\},
   \label{john1}\\
  \lefteqn{ O^T_j(\xi) \; + \; \widetilde{O}^T_{j+1}(\xi)} \nonumber \\
  & \phantom{****} = & - \; \frac{1}{j+1} \;
   \left( 1 \; - \; \frac{1}{j}\; \xi\frac{d}{d\xi} \right)
   \sum_{{\rm even}\; k = 2}^{j} \xi^k \; O^L_{j-k}(\xi) \; + \;
   \frac{1}{j+1} \left( j \; + \; \frac{1}{j} \; \xi\frac{d}{d\xi} \right)
   O^L_j(\xi)\nonumber \\
  & \phantom{****} = & \sum_{{\rm even}\; k = 0}^j \xi^k \;
   \left\{ \widetilde{C}^{k, j}_{a+} \; O^L_{j-k+1}(\xi) \; + \;
   \widetilde{C}^{k,j}_{b+} \; \xi\frac{d}{d\xi} \; O^L_{j-k+1}(\xi) \right\}.
\label{john2}\eea
Here, the Wilson coefficients are
\bea
  \widetilde{C}^{k, j}_{a-} & = & 
   (1 \, - \, \delta_{k,0}) \; \frac{j - k + 1}{j(j+1)}, \\
  \widetilde{C}^{k,j}_{b-} & = & \delta_{k,0} \; \frac{1}{j+1} \; - \;
   (1 \, - \, \delta_{k,0}) \; \frac{1}{j(j+1)}, \\
  \widetilde{C}^{k, j}_{a+} & = & \delta_{k,0} \; \frac{j}{j+1} \; - \;
   (1 \, - \, \delta_{k,0}) \; \frac{j - k}{j(j+1)}, \\
  \widetilde{C}^{k,j}_{b+} & = & \frac{1}{j(j+1)}.
\eea
One can check that, in each case, the coefficients obey
\be
  \widetilde{C}^{j+k, j} \; \sim \; \frac{1}{j^2}
\ee
in the large $j$ limit. Thus there is no problem with factorisation.

Finally, we reconstruct the factorised forms for the structure functions
$T_T$ and $T^\pm_3$ in the same way as for the simple model considered
at the end of the last section. We obtain
\bea
  T_T(x, \xi, Q^2, t) & = & - \; \frac{1}{2} \int_{-1}^1 \frac{du}{u} \;
   \frac{2u^2}{u^2 - x^2} \; q(u, \xi, t), \label{final_begin}\\
  T_L(x, \xi, Q^2, t) & = & 0, \\
  T^-_3(x, \xi, Q^2, t) & = & - \; \frac{1}{2}\int_{-1}^1 \frac{du}{u} \;
   \left\{ \left[ \frac{u^2}{u^2-x^2} \; + \; \frac{2 u^2 \xi^2}{(u^2-\xi^2)^2}
   \left( \ln\left(1 - \frac{u}{x}\right) \right. \right. \right. \nonumber \\
  & & \phantom{***} \left. - \; \left(1 - \frac{x}{\xi}\right)
   \ln\left(1 - \frac{\xi}{x}\right) \; + \; 1 \right) \nonumber \\
  & & \phantom{***} \left. - \; \frac{u^2(u^2 + \xi^2)}{(u^2 - \xi^2)^2}
   \left( 1 \; + \; \frac{x}{u}\; \ln\left(1 - \frac{u}{x}\right) \right)
   \right] q(u, \xi, t) \nonumber \\
  & & + \; \left[ \frac{u^2}{u^2-\xi^2} \left(
   \left(1 - \frac{x}{u}\right) \ln\left(1 - \frac{u}{x}\right) \; - \; 1
   \right) \right. \nonumber \\
  & & \phantom{***} \left. \left. \; - \; \frac{\xi^2}{u^2-\xi^2} \left(
   \left(1 - \frac{x}{\xi}\right) \ln\left(1 - \frac{\xi}{x}\right) \; - \; 1
   \right) \right] \xi\frac{d}{d\xi} \; q(u, \xi, t) \right\} \\
  & & \phantom{********************************} + ( x \; \ra \; -x),
   \nonumber \\
  T^+_3(x, \xi, Q^2, t) & = & - \; \frac{1}{2}\int_{-1}^1 \frac{du}{u} \;
   \left\{ \left[ \frac{2u^2\xi^2}{(u^2-\xi^2)^2} \left(
   \left(- 1 + \frac{x}{u}\right) \ln\left(1 - \frac{u}{x}\right) \; + \;
   \frac{\xi - x}{u} \; \ln\left(1 - \frac{\xi}{x}\right) \right)
   \right. \right. \nonumber \\
  & & \phantom{***}  + \; \left. \frac{\xi^2}{u^2 - \xi^2} \;
   \ln \left(1 - \frac{u}{x}\right) \right] q(u, \xi, t)
   \nonumber \\
  & & + \; \left[ \frac{\xi^2}{u^2 - \xi^2}
   \left( - \, \ln\left(1 - \frac{u}{x}\right) \; + \;
   \frac{\xi - x}{u} \; \ln\left(1 - \frac{\xi}{x}\right) \right) \right.
   \nonumber \\
  & & \phantom{***} \left. \left. \frac{u^2}{u^2 - \xi^2} \; \frac{x}{u} \;
   \ln\left(1 - \frac{u}{x}\right) \right]
   \xi\frac{d}{d\xi} \; q(u, \xi, t) \right\} \label{final_end}\\
  & & \phantom{********************************} - ( x \; \ra \; -x). \nonumber
\eea
In these expressions, the twist-2 quark SPDF is introduced such that
\be
  \int_{-1}^1 du \; u^{j-1} \; q(u, \xi, t) \; = \; O^L_j(\xi).
\ee
Notice that only its odd part w.r.t. $u$ can contribute to any of the
structure functions. Note also that, as claimed, the limit $x \ra \pm\xi$
is finite and that all singularities as $u \ra \pm\xi$ cancel out both
for the coefficients of $q(u, \xi, t)$ and those of
its derivative w.r.t. $\xi$.


\section{Summary}
\label{sec_sum}

In summary, we have demonstrated factorisation for the DVCS amplitude
at leading twist using the local OPE. We showed the three main steps
in deriving this factorisation: (1) Obtain the large $x$ with the OPE and
identify the coefficients with the moments of the spectral function;
(2) Invert the moments; (3) Reconstitute the factorised amplitude using
the dispersion relation. These steps we illustrated with a simple model
obeying the condition under which the factorisation works for the DVCS
kinematics, ie. on-shell final state photon. We also showed that all
soft singularities cancel out or are integrable, given that the SPDFs
are continuous at the end of the ERBL region.

We then considered the tree-level OPE in QCD, including twist-3 contributions.
By defining the structure functions for the amplitude suitably, we
obtained them in terms of coefficient functions and the twist-2 quark SPDF.
The coefficient functions do indeed obey the factorisation condition.


\section*{Appendix}

This section provides a summary of the main properties of Skewed Parton
Distribution Functions (SPDFs). We first define a twist-2 SPDF $f(u,\xi,t)$
and briefly note some of its properties. Then we consider its continuity
at the edges of the ERBL region $-|\xi| < u < |\xi|$. 

In a scalar theory, the (twist-2) SPDF is defined as the Fourier transform
of the off-forward matrix element of the time-ordered product of
two scalar operators,
\be
  f(u, \xi, t) \; = \; \frac{1}{2\pi} \; u \int_{-\infty}^\infty d\lambda \;
   e^{i\lambda u} \mat{p_2}{\; T(\; \phi(-\half \lambda n) \;
   \phi(\half \lambda n) \; ) \;}{p_1},
\label{spdf_def}\ee
where $n^\mu$ is such that $n^2 = 0$, $n\cdot p = 1$ and $n.\Delta = 2\xi$.
In the limit $\xi \ra 0$, this reduces to the ordinary parton distribution,
$f(u)$. This limit can be taken by $p_2 \ra p_1$.

Important properties of $f(u, \xi, t)$ are as follows:
\begin{enumerate}
 \item $f(u, \xi, t) = -f(-u, \xi, t)$. This can be seen by replacing
  $u \ra -u$ in~(\ref{spdf_def}) and changing integration variables
  $\lambda \ra -\lambda$ to compensate.
 \item $f(u, \xi, t) = f(u, -\xi, t)$. This follows from crossing symmetry,
  whereby we swap $p_1$ and $p_2$.
 \item $f(u, \xi, t)$ has support $-1 < u < 1$. To see this, we take
  moments of~(\ref{spdf_def}) as follows.
\end{enumerate}
\bea
  \lefteqn{\int_{-\infty}^\infty du \; u^{j-1} \; f(u, \xi, t)} \nonumber \\
  & = &
   \frac{1}{2\pi} \int_{-\infty}^\infty du \int_{-\infty}^\infty d\lambda \;
   \mat{p_2}{\; T(\; \phi(-\half \lambda n) \;
   \phi(\half \lambda n) \; ) \;}{p_1} \;
   \left(-i \;\frac{d}{d\lambda}\right)^j e^{i\lambda u} \nonumber \\
  & = & n_{\mu_1} \cdots n_{\mu_j} \; \mat{p_2}{ : \phi(0)
   \,i\hpart\mbox{}^{\mu_1} \, i\hpart\mbox{}^{\mu_j} \,\phi(0) : }{p_1}.
\eea
Thus, the moments are equal to matrix elements of local, composite operators.
The support properties of $f$ follow from the analyticity of the these
matrix elements. In general, if we have an analytic function g(j), and
\be
  \int_{-\infty}^\infty du \; u^{j-1} \; f(u) \; = \; a^j \; g(j),
\ee
and provided g(j) has no poles for ${\rm Re}j > c$, with $c$ a positive real
number, then
\be
  f(u) \; = \; \theta\left( 1 - \left|\frac{u}{a}\right|\right)
   \widetilde{f}\left(\frac{u}{a}\right),
\label{support}\ee
with $\widetilde{f}$ a function or distribution. For $a = 1$, we
get a function with support $-1 < u < 1$. The matrix elements
can be expected to have these properties, so that $f(u, \xi, t)$ does indeed
have support $-1 < u < 1$.

The moments of $f$ have another important property: They are even polynomials
in $\xi$ of order $j$. To see this we decompose the matrix elements into
a series of all the possible Lorentz structures, given that the matrix
elements are those of a scalar target:
\bea
  \int_{-1}^1 du \; u^{j-1} \; f(u, \xi, t) & = & n_{\mu_1} \cdots n_{\mu_j}
   \sum_{{\rm even}\; m=0}^j
   \raisebox{-0.12in}{$\stackrel{\textstyle S}{\scriptstyle \{\mu\}}$} \;
   \half \Delta^{\mu_1} \cdots \half \Delta^{\mu_m} p^{\mu_{m+1}} \cdots
   p^{\mu_j} \; \langle O^{m,j} \rangle \nonumber \\
  & = & \sum_{{\rm even}\; m=0}^j \xi^m \; \langle O^{m,j} \rangle
\eea
Thus the moments are indeed polynomials of $\xi$ of order $j$.

Let us explore now the analyticity of these polynomials.
If we assume that the reduced matrix elements
$\langle O^{m,j} \rangle$ can be treated as analytic functions of j, then
all that remains is to rearrange the sum over powers of $\xi$ by expressing
it as a difference of two infinite series:
\be
  \int_{-1}^1 du \; u^{j-1} \; f(u, \xi, t) \; = \; 
   \sum_{{\rm even}\; m=0}^\infty \xi^m \langle O_{m,j} \rangle \; - \;
   \xi^j \sum_{{\rm even}\; m=2}^\infty \xi^m \langle O^{m+j,j} \rangle.
\ee
By assumption, the first
infinite sum is an analytic function of $j$. However, the second
sum has a factor of $\xi^j$. By eq.~(\ref{support}), it then follows that
\be
  f(u, \xi, t) \; = \; \theta(1 - |u|) \; \widetilde{f}(u, \xi, t)
   \; - \; \theta\left( 1 - \left|\frac{u}{\xi}\right|\right) 
   \widetilde{f}^\prime(\frac{u}{\xi}, \xi, t).
\label{regions}\ee
Here, $\widetilde{f}$ and $\widetilde{f}^\prime$ are functions
or distributions of $u$, such that
\bea
  \int_{-1}^1 du \; u^{j-1} \;\widetilde{f}(u, \xi, t) & = &
   \sum_{{\rm even}\; m=0}^\infty \xi^m \langle O^{m,j} \rangle,\label{dglap}\\
  \int_{-1}^1 du \; u^{j-1} \;\widetilde{f}^\prime(u, \xi, t) & = &
   \sum_{{\rm even}\; m=2}^\infty \xi^m \langle O^{m+j,j} \rangle.
\eea

Eq.~(\ref{regions}) is important because it explicitly separates
two contributions to the SPDFs. The first term on the r.h.s. we shall refer
to as the DGLAP contribution since its evolution is given by a generalised
DGLAP equation. The second term, which vanishes in the forward
limit $\xi \ra 0$, shall be called the ERBL contribution, since it resembles,
in its support properties, light-cone wavefuntions which have an ERBL-type
evolution. The region $|u| < |\xi|$ region we refer to as the ERBL region.

Eq.~(\ref{regions}) enables us to see under what condition the SPDFs are
continuous at the edges of the ERBL region $|u| = |\xi|$. This
is significant because integrability of the soft singularities which
appear in the DVCS coefficient functions requires it. If
$\widetilde{f}^\prime(u/\xi, \xi, t)$ vanishes as $|u| \ra |\xi|$ then
the SPDF is indeed continuous. A sufficient condition for this to happen is
that the moments of $\widetilde{f}^\prime$ vanish as fast as $1/j^2$ for 
large $j$. Explicitly,
\be
  \sum_{{\rm even}\; m=2}^\infty \xi^m \langle O^{m+j,j} \rangle \;
   \stackrel{j \ra \infty}{\lngra} \; \frac{1}{j^2} \;\; \Rightarrow \;\;
   \widetilde{f}^\prime(u/\xi, \xi, t) \;
   \stackrel{|u| \ra |\xi|}{\lngra} 0.
\ee
To our knowledge, no proof of this behaviour as yet exists, although it
is believed to be true.


\subsection*{Acknowledgements}

The author is indebted for helpful conversations to J. Forshaw, M. McDermott
and especially G. Shore under whose PhD supervision much of this work was
done. The support of PPARC is gratefully acknowledged.


\end{document}